\documentclass[twocolumn,showpacs]{revtex4}
\usepackage{graphicx}
\usepackage{dcolumn}

\begin{document}


\title{Different particle alignments in $N \approx Z$ Ru isotopes
       studied by the shell model}

\author{M. Hasegawa$^{1}$, K. Kaneko$^{2}$, T. Mizusaki$^{3,4}$ and S. Tazaki$^{5}$}
\affiliation{
$^{1}$Laboratory of Physics, Fukuoka Dental College, Fukuoka 814-0193, Japan \\
$^{2}$Department of Physics, Kyushu Sangyo University, Fukuoka 813-8503, Japan \\
$^{3}$Institute of Natural Sciences, Senshu University, Kawasaki, Kanagawa, 214-8580, Japan \\
$^{4}$Institute for Solid State Physics, University of Tokyo, Kashiwanoha, Kashiwa,
 277-8581, Japan \\
$^{5}$Department of Applied Physics, Fukuoka University, Fukuoka 814-0180, Japan
}

\date{\today}

\begin{abstract}

  Experimentally observed heaviest $N \approx Z$ nuclei, Ru isotopes, are
 investigated by the shell model on a spherical basis with the extended
 $P+QQ$ Hamiltonian.  The energy levels of all the Ru isotopes can be
 explained by the shell model with a single set of force parameters.
  The calculations indicate enhancement of quadrupole correlations
 in the $N=Z$ nucleus $^{88}$Ru as compared with the other Ru isotopes,
 but the observed moments of inertia seem to require much more enhancement
 of quadrupole correlations in $^{88}$Ru.  It is discussed
 that the particle alignment takes place at $8^+$ in $^{90}$Ru
 but is delayed in $^{88}$Ru till $16^+$ where the simultaneous alignments
 of proton and neutron pairs take place.
 The calculations present interesting predictions for $^{89}$Ru
 that the ground state is the $1/2^-$ state and there are three $\Delta J=2$
 bands with different particle alignments including the $T=0$ $p-n$ pair
 alignment.

\end{abstract}


\maketitle

\section{Introduction}

 The so-called delay of alignment in the $N=Z$ even-even nuclei is
 observed in the $64 \le A \le 88$ region \cite{Angel,Fisher,Marg1,Marg2}
 and in the lighter nucleus $^{48}$Cr.
 This phenomenon is a sign of strong proton-neutron $(p-n)$ correlations
 in the same shell \cite{Goodman}, and the special collectivity
 in the $N=Z$ even-even nuclei suggests a strong collaboration of $p-p$,
 $n-n$ and $p-n$ correlations in the $A=4m$ nuclei with $N=Z=2m$
 which can be called the $\alpha$-like ($T=0$) $2p-2n$ correlations
 \cite{Marumori,Danos,Arima}.
 A theoretical investigation of this is challenging,
 which belongs to the study of the properties of the $p-n$ interaction
 in $N=Z$ nuclei \cite{Frauendorf,Kaneko1,Kaneko2,Sheikh,Satula,Wyss,Sun1}.
 The experimental study of heavy $N=Z$ nuclei has reached $^{88}$Ru
 \cite{Marg1,Marg2}.  The new data have revealed that there is a remarkable
 difference between neighboring even-even nuclei with $N=Z$ and $N=Z+2$
 in the $1g_{9/2}$-subshell region. The qualitative difference between
 $^{88}$Ru and $^{90}$Ru ($^{84}$Mo and $^{86}$Mo) in the backbending plots
 of the yrast bands is different from the conditions
 in lighter nuclei Zr, Sr, {\it etc.}, and casts a new light on the problem
 of the delayed alignment.
 
   A theoretical explanation of the delayed alignment for heavy $N=Z$ nuclei
 $^{84}$Mo, $^{88}$Ru {\it etc.} is presented in Refs. \cite{Marg2,Sun2}
 with the projected shell model on the deformed basis \cite{Hara,Sun}.
 The projected shell model reproduces the graphs of observed moments of
 inertia, by adopting commonly accepted deformations for those nuclei.
 The adopted deformations manifest that the deformation is larger
 for even-even $N=Z$ nuclei when compared with $N>Z$ nuclei.
 In other words, the delayed alignment in $^{88}$Ru is related to
 the large deformation.
 The study with the projected shell model \cite{Marg2,Sun1} suggested
 an enhancement of the $p-n$ quadrupole-quadrupole ($QQ$) interaction
 in the $N=Z$ nuclei.
 It is our interest to understand the structural difference between
 the $N=Z$ even-even nucleus $^{88}$Ru and neighboring isotopes
 in various aspects. In this paper, we make the study using the shell model
 calculations on the spherical basis which is free from fixing the deformation
 parameter.

   The extended $P+QQ$ model \cite{Hase1,Hase2} reproduces
 observed energy levels and $B(E2)$ in $N \approx Z$ $1f_{7/2}$-subshell
 nuclei and is capable of describing the backbending phenomena.
  It has successfully clarified characteristics of the structure of
 a heavier $N=Z$ nucleus $^{64}$Ge in the configuration space
 $(2p_{3/2},1f_{5/2},2p_{1/2},1g_{9/2})^8$ in a recent shell model
 calculation \cite{Kaneko}.  The heaviest $N=Z$ nucleus experimentally
 observed, $^{88}$Ru, which is expected to have the approximate
 configuration $(2p_{3/2},1f_{5/2},2p_{1/2},1g_{9/2})^{-12}$,
 is a good target to study the delayed alignment using the shell model
 calculation.  The success in $^{64}$Ge suggests that the extended $P+QQ$
 model provides a reliable interaction for the study of the heavy
 $N \approx Z$ nuclei.  We carry out shell model calculations
 using the extended $P+QQ$ model with a single set of force parameters
 fixed for $^{88}$Ru and heavier Ru isotopes.
 The calculations, which are carried out with the calculation
 code \cite{Mizusaki}, have huge dimensions (maximum dimension is
 $165\times 10^6$ for $^{88}$Ru) and can be regarded as realistic ones.
 We investigate the structure of Ru isotopes and examine
 whether the difference between $^{88}$Ru and $^{90}$Ru is reproduced
 or not by the spherical shell model, in Section III.
 The present shell model predicts interesting features of the odd-$A$
 isotope $^{89}$Ru between $^{88}$Ru and $^{90}$Ru.
 The prediction for $^{89}$Ru is shown in Section IV.

   Since the delayed alignment in $^{88}$Ru seems to be related to
 the strong quadrupole correlations and the large quadrupole deformation
 \cite{Marg2,Sun1},
 we pay our attention to the roles of the $QQ$ force 
 which induces the quadrupole correlations and deformation. 
 It is interesting to see the competition between the like-nucleon
 ($p-p$ and $n-n$) interaction and $p-n$ interaction of the $QQ$ force.
  We also examine a possible contribution of the isovector $QQ$ force
 to the properties of Ru isotopes.

\section{The model Hamiltonian}

The extended $P+QQ$ Hamiltonian is given by
\begin{eqnarray}
 H & = & H_{\rm sp} + H_{\rm mc}
        + H_{P_0} + H_{P_2} + H^{\tau =0}_{QQ} + H^{\tau =0}_{OO}  \nonumber \\
   & = & \sum_{\alpha} \varepsilon_a c_\alpha^\dag c_\alpha + H_{\rm mc}
        -\sum_{J=0,2} \frac{1}{2} g_J \sum_{M\kappa} P^\dag_{JM1\kappa} P_{JM1\kappa}
          \nonumber \\
   & - & \frac{1}{2} \frac{\chi^0_2}{b^4} \sum_M :Q^\dag_{2M} Q_{2M}: 
         - \frac{1}{2} \frac{\chi^0_3}{b^6} \sum_M :O^\dag_{3M} O_{3M}:, \label{eq:1}
\end{eqnarray}
 where $\varepsilon_a$ is a single-particle energy, $H_{\rm mc}$ denotes
 the monopole corrections, $P_{JMT\kappa}$ is the pair operator with angular
 momentum $J$ and isospin $T$, and $Q_{2M}$ ($O_{3M}$)  is the isoscalar
  quadrupole (octupole) operator (see Ref. \cite{Hase2}).
 The force strengths $\chi^0_2$ and $\chi^0_3$ are defined so as to have
 the dimension of energy.  Following Ref. \cite{Kaneko}, we adopt the model space
 $(2p_{3/2}, 1f_{5/2},2p_{1/2},1g_{9/2})$ and introduce the isoscalar
 octupole-octupole force $H^{\tau =0}_{OO}$.
 Note that the Hamiltonian is isospin-invariant and includes
 the $p-n$ pairing forces in addition to the $p-n$ $QQ$ force.
 
   The isoscalar $QQ$ force $H^{\tau =0}_{QQ}$ can be divided into three parts,
 $p-p$, $n-n$ and $p-n$ parts.  We shall use the notations
 $\chi^0_{2pp}$, $\chi^0_{2nn}$ and $\chi^0_{2pn}$ for their force strengths.
 In terms of the pair operators $P_{JMT\kappa}$, the isoscalar $QQ$ force
 is expressed as
\begin{eqnarray}
 H^{\tau =0}_{QQ} & = & - \frac{1}{2} \sum_\kappa \frac{x^1_\kappa}{b^4}
                  \sum_{JM} W_{JT=1} P^\dag_{JM1\kappa} P_{JM1\kappa}  \nonumber \\
       & {} & - \frac{1}{2} \frac{x^0_{\kappa=0}}{b^4}
                  \sum_{JM} W_{JT=0} P^\dag_{JM00} P_{JM00},   \label{eq:2}
\end{eqnarray}
 where $x^1_\kappa = x^0_{\kappa=0} = \chi^0_2$.
 The symbol $W_{JT}$ proportional to the Racah coefficient really has four
 subscripts related to the four orbits of
 $c_{\alpha}^{\dag} c_{\beta}^{\dag} c_\delta c_\gamma $.
 The first line of Eq. (\ref{eq:2}) brings about the isovector pairing
 interactions, where the strengths
 $x^1_{\kappa=1}$, $x^1_{\kappa=-1}$ and $x^1_{\kappa=0}$
 stand for the $n-n$, $p-p$ and $p-n$ interactions.
 The second line of Eq. (\ref{eq:2}) brings about the isoscalar $p-n$ pairing
 interactions. 
 The $p-n$ part of $H^{\tau =0}_{QQ}$ is enhanced by enlarging the force
 strength $\chi^0_{2pn}$ ($x^1_{\kappa=0}$ and $x^0_{\kappa=0}$)
 in Refs. \cite{Marg2,Sun1,Kaneko}.
  The Hamiltonian ceases to be isospin-invariant, with the isospin
 not being a good quantum number there.

  There is a possibility of the isovector $QQ$ force contributing to
 the collective motion in the heavy $N \approx Z$ nuclei.
 The isovector $QQ$ force $H^{\tau =1}_{QQ}$ with the force strength
 $\chi^1_2$ is also rewritten in the same form as Eq. (\ref{eq:2})
 with the relations $x^1_\kappa = \chi^1_2$ and
  $x^0_{\kappa=0} = -3 \chi^1_2$.
 We can write the sum of $H^{\tau =0}_{QQ}$ and $H^{\tau =1}_{QQ}$
 in the same form as Eq. (\ref{eq:2}), where
  $x^1_\kappa = \chi^0_2 + \chi^1_2$ and
  $x^0_{\kappa=0} = \chi^0_2 -3 \chi^1_2$.
 If we consider a restricted sum of $H^{\tau =0}_{QQ}$ and $H^{\tau =1}_{QQ}$
 with the following combination of the interaction strengths
\begin{equation}
  \chi^0_2 = (1+\alpha) x , \quad \chi^1_2 = - \alpha x, \label{eq:3}
\end{equation}
the $QQ$ force is written as
\begin{eqnarray}
 H^{\tau =0}_{QQ} & + & H^{\tau =1}_{QQ}                         \nonumber \\
 & = & - \frac{1}{2} \frac{x}{b^4} \sum_\kappa 
                  \sum_{JM} W_{JT=1} P^\dag_{JM1\kappa} P_{JM1\kappa}  \nonumber \\
 & {}& - \frac{1}{2} \frac{(1+4\alpha)x}{b^4}
                  \sum_{JM} W_{JT=0} P^\dag_{JM00} P_{JM00}.   \label{eq:4}
\end{eqnarray}
 By changing the mixing parameter $\alpha$, we can enhance the $p-n$ part of
 the $QQ$ force, which corresponds to the isoscalar $p-n$ pairing interactions
 in the second line of Eq. (\ref{eq:4}), without violating the isospin invariance
 of the Hamiltonian.

\section{Difference between $^{88}$Ru and $^{90}$Ru}

\begin{figure}[b]
\includegraphics[width=8cm,height=8cm]{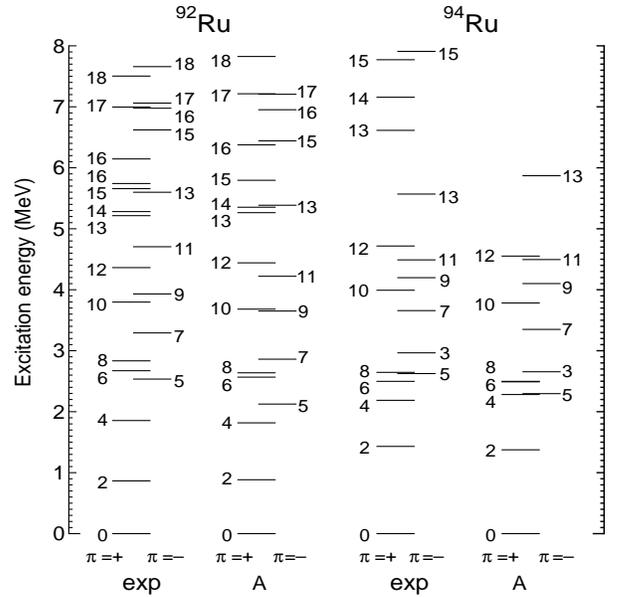}
  \caption{Energy levels of $^{92}$Ru and $^{94}$Ru.
           The label ``A" stands for the energy levels calculated
           with the parameter set (5) and ``exp" for the observed ones.}
  \label{fig1}
\end{figure}

   Using the extended $P+QQ$ Hamiltonian (\ref{eq:1}), we carried out
 shell model calculations in the hole space
 $(1g^h_{9/2},2p^h_{1/2},1f^h_{5/2},2p^h_{3/2})$ with the calculation
 code \cite{Mizusaki}.
 The single-hole energies $\varepsilon^h_a$ depend on $H_{\rm mc}$ and
 the force strengths as well as $\varepsilon_a$ through the hole
 transformation.
 We treated the hole energies $\varepsilon^h_a$ as parameters instead of
 the single-particle energies $\varepsilon_a$.
 We tried various combinations of the parameters $\varepsilon^h_a$,
 $H_{\rm mc}$, $g_0$, $g_2$, $\chi^0_2$ and $\chi^0_3$, and determined these
 parameters so as to reproduce overall energy levels of the Ru isotopes.  
 The adopted parameters are 
\begin{eqnarray}
 & {} & \varepsilon^h_{9/2} = 0.0, \quad \varepsilon^h_{1/2} = 1.1,
   \quad \varepsilon^h_{5/2} = 5.5, \quad \varepsilon^h_{3/2} = 6.0,  \nonumber \\
 & {} &  g_0 = 0.26 (92/A), \quad g_2 = 0.12 (92/A)^{5/3},            \nonumber \\
 & {} &  \chi^0_2 = 0.26 (92/A)^{5/3}, \quad \chi^0_3 = 0.04 (92/A)^2
    \mbox{ in MeV},                                              \label{eq:5}
\end{eqnarray}
 and $H_{\rm mc}$ is fixed at zero (the $J$-independent isoscalar monopole
 term is not determined, because we do not deal with the binding
 energy in this paper).  Changing the monopole corrections
 $H_{\rm mc}$ does not significantly improve the energy levels.
 The relative position of $\varepsilon^h_{9/2}$ and $\varepsilon^h_{1/2}$
 is responsible to that of the positive and negative parity states.
 In our trials, the values of $\varepsilon^h_{5/2}$ and $\varepsilon^h_{3/2}$
 listed in (\ref{eq:5}) are best and the exchange of the two values does not
 improve the energy levels.  The hole levels $1f^h_{5/2}$ and $2p^h_{3/2}$
 seem to lie far from $2p^h_{1/2}$.
 This is the reason why the subspace $(2p_{1/2},1g_{9/2})$ works well
 for $A>86$ nuclei in Refs. \cite{Serduke,Herndl}.  The force strengths
 $g_0$, $g_2$, $\chi^0_2$ and $\chi^0_3$ adopted are similar to those
 used in the study of $^{64}$Ge \cite{Kaneko}.

\begin{figure}[h]
\includegraphics[width=8cm,height=8cm]{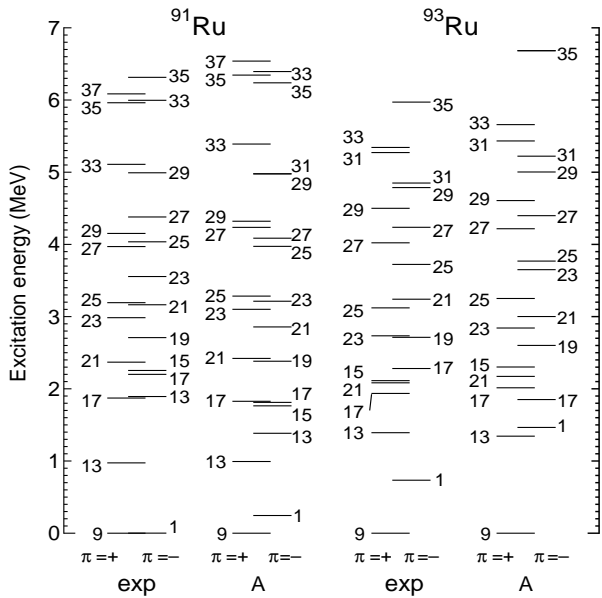}
  \caption{Calculated and observed energy levels of $^{91}$Ru and $^{93}$Ru.
           The spin of each state is denoted by the double number $2J$.}
  \label{fig2}
\end{figure}

    The parameter set (\ref{eq:5}) reproduces well the energy levels
 (the patterns and order of the positive- and negative-parity levels)
 of Ru isotopes, not only the even-$A$ nuclei $^{92}$Ru and $^{94}$Ru
 but also the odd-$A$ nuclei $^{91}$Ru and $^{93}$Ru as shown in Figs.
 \ref{fig1} and \ref{fig2}. 
 The agreement between theory and experiment for the odd-parity state
 is worse than that for the even-parity states.  The calculation, however,
 reproduces the observed energies within the error 0.8 MeV.

\subsection{Dependence on the $QQ$ force strengths}

\begin{figure}
\includegraphics[width=7cm,height=7cm]{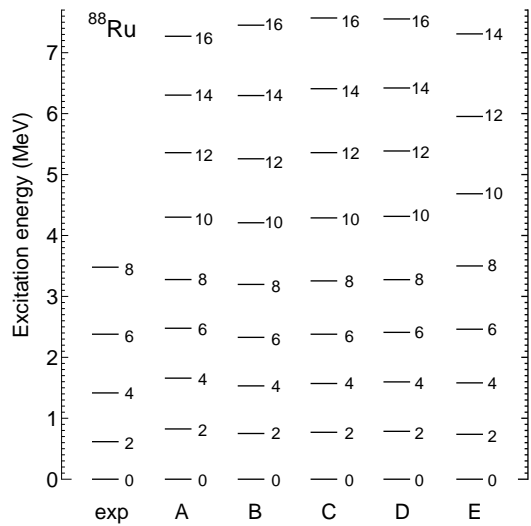}
  \caption{Comparison of calculated energy levels with observed ones for $^{88}$Ru.
          The calculated results are obtained with the different strengths
          of the $QQ$ force, A, B, C, D and E.} 
  \label{fig3}
\end{figure}

\begin{figure}[b]
\includegraphics[width=7cm,height=7cm]{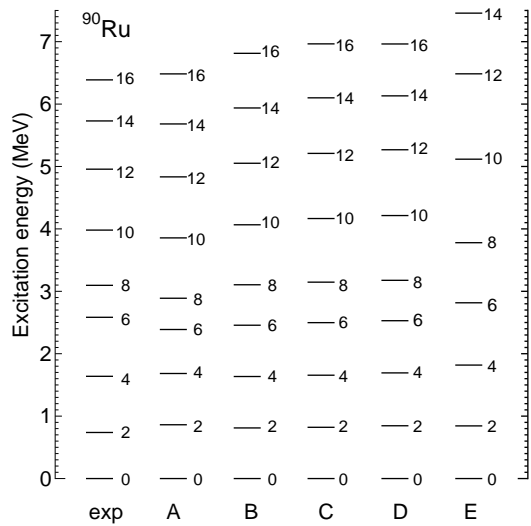}
  \caption{Comparison of calculated energy levels with observed ones
           for $^{90}$Ru.  The calculated results are obtained with
           the different strengths of the $QQ$ force, A, B, C, D and E.} 
  \label{fig4}
\end{figure}

\begin{figure}
\includegraphics[width=7cm,height=7cm]{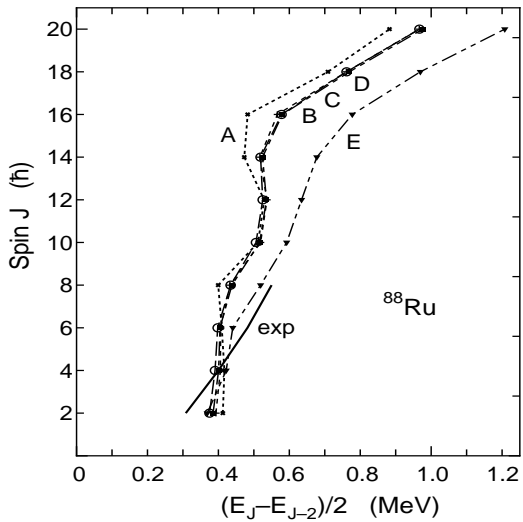}
  \caption{The $J-\omega$ graph of Fig. 1.}
  \label{fig5}
\end{figure}

\begin{figure}[h]
\includegraphics[width=5.0cm,height=7cm]{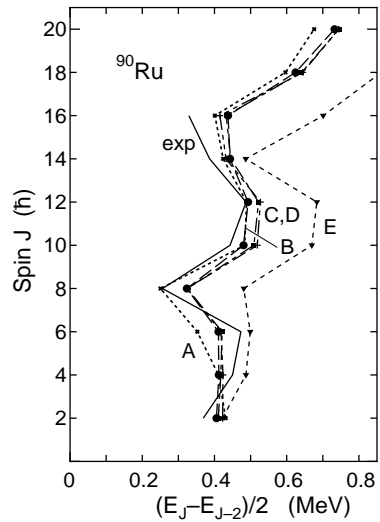}
  \caption{The $J-\omega$ graph of Fig. 2.}
  \label{fig6}
\end{figure}

   The energy levels obtained for $^{88}$Ru and $^{90}$Ru, which are shown
 in the column A of Figs. \ref{fig3} and \ref{fig4}, are consistent
 with the observed ones.
  The parameter set A describes the difference between $^{88}$Ru and $^{90}$Ru
 in the backbending plot (we call it ``$J-\omega$ graph") as shown in 
 Figs. \ref{fig5} and \ref{fig6}.
 The calculation reproduces the sharp backbending at $J=2j-1=8$ observed
 in $^{90}$Ru and shows no clear backbending in low-spin states of $^{88}$Ru
 in agreement with the experiment.
 
 The backbending plots, however, reveal insufficiency for the most collective
 low-lying states.  The results A do not reproduce the slope of
 the $J-\omega$ graph up to $J=8$ for $^{88}$Ru and  up to $J=6$ for $^{90}$Ru.
  The slopes of the $J-\omega$ graphs for the collective bands are considerably
 affected by the strength $\chi^0_2$ of the $QQ$ force $H^{\tau =0}_{QQ}$
 above all other force strengths.  This is naturally understood,
 because the moment of inertia of a rotational band depends on the magnitude
 of deformation and the $QQ$ force drives the quadrupole deformation.
   Let us try to improve the $J-\omega$ graphs for $^{88}$Ru and $^{90}$Ru
 by readjusting the $QQ$ force strength.
 
    We first strengthen the $p-n$ $QQ$ interaction by adding the isovector
 $QQ$ force $H^{\tau =1}_{QQ}$ in the form (\ref{eq:4}) so as to conserve
 the isospin invariance.
 The results obtained with the mixing parameter $\alpha =0.125$
 (see Eq. (\ref{eq:3})) are shown by the notation B in
 Figs. \ref{fig3} - \ref{fig6}.  The $J-\omega$ graph is
 improved for $^{90}$Ru.  For $^{88}$Ru, the result B removes the slight
 backbending at $J=8$ of the result A.
 The parameter set B reproduces quite well the overall energy levels of
 the Ru isotopes $^{90}$Ru, $^{91}$Ru, $^{92}$Ru, $^{93}$Ru and $^{94}$Ru.
   For the high-spin states, however, the parameter set A is better than B.
 The change from A to B pushes up the high-spin levels higher
 as the spin $J$ increases.   Since the configuration $(1g^h_{9/2})^m$ is 
 dominant in the high-spin states, the inadequacy for the high-spin states
 suggests that the enhanced $p-n$ $QQ$ force strength ($x^0_{\kappa =0} = 1.5x$)
 is too strong for the $1g_{9/2}$ subshell.
 The remaining deviation of the calculated $J-\omega$ graph from
 the experimental one for $^{88}$Ru indicates room for improvement
 in the model space and in the interactions of our model.

   Results similar to those of B are obtained by strengthening the $p-n$ part
 of the isoscalar $QQ$ force $H^{\tau =0}_{QQ}$ (by enlarging $\chi^0_{2pn}$).
 The results of $\chi^0_{2pn}=1.25\chi^0_{2pp}$ are shown by the notation C
 in Figs. \ref{fig3} - \ref{fig6}.  (Note that the $p-n$ force strength
 $\chi^0_{2pn}=1.3\chi^0_{2pp}$ is used by Sun {\it et al.} \cite{Marg2}
 to increase the particle alignment frequency for $^{88}$Ru and $^{84}$Mo.)
 Although the high-spin levels of C are pushed up a little higher as compared with
 those of B, the parameter sets B and C yield similar results, not only
 for the energy levels but also for $B(E2)$ values and the quadrupole moment $Q$
 with respect to the yrast bands of $^{88}$Ru and $^{90}$Ru as shown later on.
 We also tried to strengthen all of the $p-p$, $n-n$ and $p-n$ $QQ$ interactions.
 The enlargement of $\chi^0_2$ to $1.1 \times \chi^0_2$ yields results
 similar to those of B and C.  The results are denoted by D in Figs.
 \ref{fig3} - \ref{fig6}.
 Within the small increase of the $QQ$ force, we have found no evidence
 that the $p-n$ QQ interaction is stronger than the $p-p$ and $n-n$ $QQ$
 interactions, and there is no choice between the isospin-variant and -invariant
 enhancements of the $p-n$ $QQ$ interaction, as well.
 
 It should be noted here that the strength of the $p-n$ interaction does not
 directly correspond to the strength of the $p-n$ correlations.
 According to the single-$j$ shell calculation with the extended $P+QQ$ force
 \cite{Hase3}, the $p-n$ correlation energy becomes largest at $N=Z$
 in nuclei with the same $Z$ even though the same $p-n$ interaction is used
 for those nuclei, while the $p-p$ and $n-n$ correlation energy does not
 show such a specific feature.

    In Fig. \ref{fig5}, the discrepancy between the calculated moments
 of inertia (B, C and D) and observed ones is still large for $^{88}$Ru.
  The calculations B, C and D cannot sufficiently reproduce
 the observed large angular frequency at $J=8$.
 If we want to obtain a better slope of the $J-\omega$ graph for $^{88}$Ru,
 we must enhance the $QQ$ force strength $\chi^0_2$ much more. 
  The slope of the $J-\omega$ graph observed in $^{88}$Ru cannot be well
 reproduced even by strengthening the $p-n$ $QQ$ interaction further
 in the way B or C.
  Results obtained with the isoscalar $QQ$ force $H^{\tau =0}_{QQ}$
 strengthened by 1.5 ($1.5 \times \chi^0_2$) are denoted by E
 in Figs. \ref{fig3} - \ref{fig6}.
 The calculation E reproduces the large angular frequency and the slope
 (moment of inertia) of the $J-\omega$ graph for $^{88}$Ru
 but yields rather bad results for $^{90}$Ru.
 This suggests that the collectivity of the quadrupole correlations
 is different between $^{88}$Ru and $^{90}$Ru.
 The observed moment of inertia and large angular frequency indicate
 a stable rotation of $^{88}$Ru, while the energy levels and the sharp
 backbending at $8^+$ reveal a deviation from the rotation in $^{90}$Ru.
 Our spherical shell model calculation predicts that clear backbending
 does not occur up to $J=14$ in $^{88}$Ru, as predicted by the projected
 shell model \cite{Marg2,Sun2}.
 The present results indicate a special enhancement of the quadrupole
 correlations in $^{88}$Ru in contrast to the other Ru isotopes with $N>Z$.
 This is consistent with the results obtained by the projected shell model
 \cite{Marg2,Sun2}, in which the deformation parameter is fixed to be 0.23
 for $^{88}$Ru and 0.16 for $^{90}$Ru 
  (the former is 1.4 times as large as the latter).
 Our spherical shell model requires the enhancement of the $QQ$ force
 instead of the enlargement of the deformation for the $N=Z$ nucleus $^{88}$Ru.
 This suggests that the present model space
 $(2p_{3/2},1f_{5/2},2p_{1/2},1g_{9/2})$ may not be sufficient and
 the lower orbit $1f_{7/2}$ or the upper one $2d_{5/2}$ should be
 included possibly.

\subsection{Difference between $^{88}$Ru and $^{90}$Ru in structure}

\begin{table}[b]
\caption{Expectation values of proton (neutron) numbers
         in the respective orbits for the yrast states of $^{88}$Ru
         calculated with the different strengths of the $QQ$ force,
         A and B.}
\begin{tabular}{c|cccc}   \hline
    \multicolumn{5}{c}{cal. A} \\
   $J$ & $p_{3/2}$ & $f_{5/2}$ & $p_{1/2}$ & $g_{9/2}$  \\ \hline
    0  &    3.87   &    5.74   &    0.73   &    5.66    \\
    2  &    3.85   &    5.72   &    0.68   &    5.75    \\
    4  &    3.84   &    5.71   &    0.62   &    5.84    \\
    6  &    3.82   &    5.68   &    0.59   &    5.90    \\
    8  &    3.83   &    5.69   &    0.66   &    5.86    \\
   10  &    3.82   &    5.67   &    0.61   &    5.89    \\
   12  &    3.81   &    5.66   &    0.57   &    5.96    \\
   14  &    3.82   &    5.67   &    0.57   &    5.94    \\
   16  &    3.85   &    5.72   &    0.71   &    5.71    \\
   18  &    3.86   &    5.73   &    0.78   &    5.63    \\
   20  &    3.98   &    5.97   &    0.91   &    4.13    \\ \hline
  
    \multicolumn{5}{c}{cal. B} \\
   $J$ & $p_{3/2}$ & $f_{5/2}$ & $p_{1/2}$ & $g_{9/2}$  \\ \hline
    0  &    3.80   &    5.65   &    0.65   &    5.89    \\
    2  &    3.79   &    5.63   &    0.66   &    5.92    \\
    4  &    3.78   &    5.62   &    0.65   &    5.95    \\
    6  &    3.77   &    5.60   &    0.66   &    5.96    \\
    8  &    3.77   &    5.60   &    0.66   &    5.96    \\
   10  &    3.77   &    5.59   &    0.67   &    5.98    \\
   12  &    3.76   &    5.58   &    0.66   &    5.99    \\
   14  &    3.77   &    5.59   &    0.65   &    5.99    \\
   16  &    3.78   &    5.60   &    0.65   &    5.97    \\
   18  &    3.78   &    5.60   &    0.65   &    5.97    \\
   20  &    3.78   &    5.59   &    0.65   &    5.98    \\ \hline
\end{tabular}
\label{table1}
\end{table}

  The backbending at $J=2j-1=8$ in the $1g_{9/2}$-subshell nucleus $^{90}$Ru
 is contrast to no backbending at $J=2j-1=6$ in the $1f_{7/2}$-subshell
 nucleus $^{50}$Cr, while the resistance to the backbending is common to
 the $N=Z$ nuclei $^{88}$Ru and $^{48}$Cr. 
  Let us discuss the difference between $^{88}$Ru and $^{90}$Ru in the
 structure which appears in the $J-\omega$ graphs of Figs. \ref{fig5} and
 \ref{fig6}.  In Tables \ref{table1} and \ref{table2},
 we tabulate the expectation values of proton and neutron numbers
 $\langle n_a \rangle$ in the respective orbits for the yrast states of
 $^{88}$Ru and $^{90}$Ru.  The tables show that more protons jump up
 from the $pf$ subshell to the $1g_{9/2}$ one in $^{88}$Ru than in $^{90}$Ru,
 and the same is true for neutrons if the extra neutron pair is subtracted
 from the neutron number $\langle n_{g9/2} \rangle$ for $^{90}$Ru.
 Two things are characteristic in the $N \approx Z$ Ru isotopes, 
 which is different from the situation of the $N \approx Z$ Cr isotopes
 in the $1f_{7/2}$ subshell.
  First, the $1g_{9/2}$ subshell where the Fermi level lies is just above
 the $pf$ subshell and there is a considerably large degree of freedom
 for $1g_{9/2}$. 
  Secondly, the two subshells have opposite parities.
 These conditions permit only nucleon pairs jumping up to $1g_{9/2}$ and
 induce strong $p-p$, $n-n$ and $p-n$ correlations in $1g_{9/2}$.
 We can suppose that the collaboration of the $p-p$, $n-n$ and $p-n$
 correlations results in the $\alpha$-like $2p-2n$ correlations
 especially in the $1g_{9/2}$ subshell, in the $N=Z$ nucleus $^{88}$Ru
 where the $p-n$ correlations are enhanced.

\begin{table}[b]
\caption{Expectation values of proton and neutron numbers
         in the respective orbits for the yrast states of $^{90}$Ru
         calculated with the different strengths of the $QQ$ force,
         A and B.}
\begin{tabular}{c|cccc|cccc}   \hline
    \multicolumn{9}{c}{cal. A} \\
       & \multicolumn{4}{c}{proton} & \multicolumn{4}{|c}{neutron} \\
   $J$ & $p_{3/2}$ & $f_{5/2}$ & $p_{1/2}$ & $g_{9/2}$
       & $p_{3/2}$ & $f_{5/2}$ & $p_{1/2}$ & $g_{9/2}$  \\ \hline
  0  & 3.94 & 5.87 & 1.08 & 5.11 & 3.99 & 5.96 & 1.84 & 6.21 \\
  2  & 3.93 & 5.86 & 1.00 & 5.20 & 3.99 & 5.97 & 1.88 & 6.16 \\
  4  & 3.93 & 5.87 & 1.01 & 5.19 & 3.99 & 5.98 & 1.94 & 6.09 \\
  6  & 3.96 & 5.90 & 1.41 & 4.73 & 3.99 & 5.98 & 1.94 & 6.08 \\
  8  & 3.97 & 5.93 & 1.64 & 4.46 & 3.99 & 5.97 & 1.88 & 6.16 \\
 10  & 3.98 & 5.95 & 1.77 & 4.30 & 3.99 & 5.98 & 1.92 & 6.12 \\
 12  & 3.98 & 5.97 & 1.89 & 4.16 & 3.99 & 5.98 & 1.95 & 6.07 \\
 14  & 3.98 & 5.96 & 1.81 & 4.25 & 3.99 & 5.99 & 1.96 & 6.06 \\
 16  & 3.97 & 5.95 & 1.64 & 4.44 & 3.99 & 5.99 & 1.97 & 6.05 \\
 18  & 3.98 & 5.96 & 1.74 & 4.32 & 3.99 & 5.99 & 1.98 & 6.03 \\
 20  & 3.99 & 5.98 & 1.88 & 4.15 & 3.99 & 5.99 & 1.98 & 6.03 \\ \hline

    \multicolumn{9}{c}{cal. B} \\
       & \multicolumn{4}{c}{proton} & \multicolumn{4}{|c}{neutron} \\
   $J$ & $p_{3/2}$ & $f_{5/2}$ & $p_{1/2}$ & $g_{9/2}$
       & $p_{3/2}$ & $f_{5/2}$ & $p_{1/2}$ & $g_{9/2}$  \\ \hline
  0  & 3.91 & 5.83 & 0.86 & 5.39 & 3.98 & 5.94 & 1.79 & 6.29 \\
  2  & 3.90 & 5.82 & 0.78 & 5.49 & 3.98 & 5.95 & 1.84 & 6.23 \\
  4  & 3.90 & 5.82 & 0.76 & 5.52 & 3.98 & 5.97 & 1.91 & 6.14 \\
  6  & 3.93 & 5.86 & 1.12 & 5.09 & 3.99 & 5.98 & 1.95 & 6.08 \\
  8  & 3.97 & 5.92 & 1.60 & 4.51 & 3.98 & 5.97 & 1.88 & 6.17 \\
 10  & 3.98 & 5.95 & 1.78 & 4.30 & 3.99 & 5.97 & 1.91 & 6.13 \\
 12  & 3.98 & 5.97 & 1.88 & 4.17 & 3.99 & 5.98 & 1.96 & 6.07 \\
 14  & 3.97 & 5.95 & 1.73 & 4.34 & 3.99 & 5.99 & 1.97 & 6.05 \\
 16  & 3.96 & 5.93 & 1.55 & 4.56 & 3.99 & 5.99 & 1.97 & 6.05 \\
 18  & 3.97 & 5.94 & 1.68 & 4.40 & 3.99 & 5.99 & 1.98 & 6.03 \\
 20  & 3.99 & 5.97 & 1.88 & 4.16 & 3.99 & 5.99 & 1.98 & 6.03 \\ \hline
\end{tabular}
\label{table2}
\end{table}

 The sharp backbending at $8^+$ in Fig. \ref{fig6} for $^{90}$Ru
 coincides with the increase of $\langle n_{g9/2} \rangle$
 and decrease of $\langle n_{p1/2} \rangle$ for neutron (and their
 decrease and increase for proton) at $J=8$ in Table \ref{table2}.
 This change is explained by the alignment of a neutron pair in $1g_{9/2}$.
 The results A and B in Table \ref{table2} suggest the following explanation:
 There is an extra neutron pair which cannot form a $T=0$ 2p-2n quartet
 \cite{Danos,Arima,Hase4} in $^{90}$Ru.  The extra neutron pair has a dominant
 probability to be a pair with $J=0$ and $T=1$, and contributes to
 the collective $2p-2n$ correlations through the exchange with a neutron
 pair in the quartets.  The excitation till $6^+$ owes to the motion
 of the quartets.  At $8^+$, the extra neutron pair aligns the angular
 momentum to be $J=9/2+7/2$ ($J=8$, $T=1$) in $1g_{9/2}$ and breaks away
 from the collective $2p-2n$ correlations, which increases the $1g_{9/2}$
 neutron number. The weakened $2p-2n$ correlations somewhat hinder
 proton pairs jumping up to the $1g_{9/2}$ subshell from the $pf$ subshell,
 which decreases the $1g_{9/2}$ proton number.
   The result A for $^{88}$Ru in Table \ref{table1} shows a similar sign
 at $J=8$, but the observed $J-\omega$ graph denies
 such a pair alignment in $^{88}$Ru.  The calculated results B, C, D
 and E which are better for the very collective low-lying states sweep away
 the sign of a structural change in $\langle n_a \rangle$ (the result B
 is shown in the lower part of Table \ref{table1}).
 By combining our result for $^{88}$Ru with the $J-\omega$ graphs
 experimentally observed in other $N=Z$ nuclei, we can say
 that the one-pair alignment is hindered in the $N=Z$ nuclei due to the
 strong $2p-2n$ correlations.
 This may be the reason for the durable increase of angular frequency
 in the $N=Z$ nuclei.
 
   Instead, Fig. \ref{fig5}, in which the monotonous slope after $J=16$
 stands out, suggests a structure change at $16^+$ in $^{88}$Ru.
 The projected shell model \cite{Marg2,Sun2} also predicts a backbending
 at $J=16$ for $^{88}$Ru.  Moreover, the calculated result A in
 Table \ref{table1} shows the increase of $\langle n_{f5/2} \rangle$ and
 $\langle n_{p1/2} \rangle$, and the decrease of $\langle n_{g9/2} \rangle$
 at $J=16$.  The same sign remains slightly in $\langle n_{g9/2} \rangle$
 of the result B and the sign disappears for the strong $QQ$ force E.
 As mentioned above, however, the enhanced $QQ$ force of B, C, D and E is
 more or less too strong for the high-spin states.   We can expect
 that the structural change at $J=16$ will be observed in $^{88}$Ru.
 This structural change seems to be caused by the simultaneous alignments of
 proton and neutron pairs at $J=2 \times (9/2+7/2)$, since the strong $2p-2n$
 correlations resist the single alignment of proton or neutron pair.
  The analysis in Ref. \cite{Hara2}, which predicts the simultaneous
 alignments of proton and neutron pairs at $J=2 \times (7/2+5/2)$
 without backbending due to the one-pair alignment in the $1f_{7/2}$ $N=Z$
 nucleus $^{48}$Cr, supports our conjecture for the $1g_{9/2}$ $N=Z$ nucleus
 $^{88}$Ru. This conjecture is also supported by the backbending toward $J=16$
 observed in $^{90}$Ru.  In Table \ref{table2}, the increase of proton
 number $\langle n^\pi_{g9/2} \rangle$ at $J=16$ in the results A and B
 for $^{90}$Ru suggests the proton-pair alignment in $1g_{9/2}$ in addition to
 the neutron-pair alignment at $J=8$.
 
\subsection{Effect of the $2d_{5/2}$ orbit}

\begin{figure}[b]
\includegraphics[width=7.4cm,height=7.4cm]{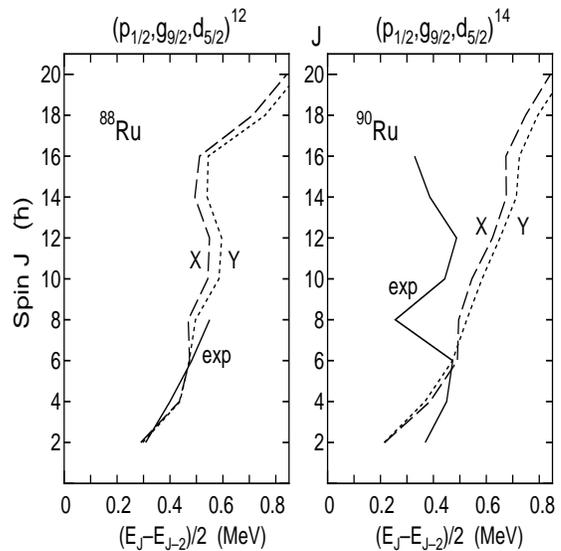}
  \caption{The $J-\omega$ graphs obtained in the model space
           $(2p_{1/2},1g_{9/2},2d_{5/2})$ for $^{88}$Ru and $^{90}$Ru,
           compared with the experimental ones.}
  \label{fig7}
\end{figure}

   The large deformation on the deformed basis can be interpreted
 by the mixing of a large number of spherical single-particle orbits.
  The expansion of the configuration space instead of the enhancement of
 the $QQ$ force is effective in our spherical shell model.
  The $2d_{5/2}$ orbit could contribute to the quadrupole
 correlations, because it strongly couples with the $1g_{9/2}$ orbit
 which plays a leading role in the Ru isotopes, through the large matrix
 element $\langle 1g_{9/2} || Q || 2d_{5/2} \rangle$.
 Adding the $2d_{5/2}$ orbit to the model space
 $(2p_{3/2},1f_{5/2},2p_{1/2},1g_{9/2})$,
 unfortunately, makes the number of the shell model basis states too huge.
 Instead of this, let us examine the contribution of the $2d_{5/2}$ orbit
 within the truncated space $(2p_{1/2},1g_{9/2},2d_{5/2})$.
 This space does not cause the spurious motion of the center of mass,
 and is expected to work well as the truncated space $(1f_{7/2},2p_{3/2})$
 without $(2p_{1/2},1f_{5/2})$ can explain the main features of
 the $f_{7/2}$-subshell nuclei \cite{Martine,Zuker} because of
 the large matrix element $\langle 1f_{7/2} || Q || 2p_{3/2} \rangle$ 
  (the $Q$ matrix element becomes large when $\Delta l= \Delta j =2$).
 We carried out the shell model calculations using the single-particle energies
 $\varepsilon_{1/2}=0.0$, $\varepsilon_{9/2}=1.0$ and $\varepsilon_{5/2}=6.0$ 
 in MeV.  The inclusion of $2d_{5/2}$ allows us to use weaker force
 strengths than those in Eq. (\ref{eq:5}).  We replaced the $A$-dependence
 $(92/A)^x$ with $(88/A)^x$ for $g_0$, $g_2$, $\chi^0_2$ and $\chi^0_3$
 in Eq. (\ref{eq:5}).
 The results for $^{88}$Ru and $^{90}$Ru are shown by the dash lines (X)
 in Fig. \ref{fig7}.
 
  In Fig. \ref{fig7}, the calculated $J-\omega$ graph agrees well
 with the experimental one observed for the low-lying collective states
 of $^{88}$Ru. The agreement is better than those of B, C and D
 in Fig. \ref{fig5}. (It is notable that Fig. \ref{fig7} also predicts
 the alignment at $J=16$ in $^{88}$Ru.)
 This suggests that adding the $2d_{5/2}$ orbit
 to $(2p_{3/2},1f_{5/2},2p_{1/2},1g_{9/2})$ with adjusted force strengths
 probably improves the calculated results for $^{88}$Ru.
 The collectivity in the expanded space could enhance the quadrupole
 correlations without strengthening the $QQ$ force. 
 Since the large deformation causes a large admixture of the $2d_{5/2}$
 spherical orbit in the deformed Nilsson basis states, the present result
 is consistent with the prediction of Ref. \cite{Marg2,Sun2}
 that the deformation of the $N=Z$ nucleus $^{88}$Ru is large (0.23).
   In contrast to this, the inclusion of $2d_{5/2}$
 ruins the $J-\omega$ graph for $^{90}$Ru as shown in Fig. \ref{fig7}.
 The energy levels obtained do not display the sharp backbending at $J=8$
 but look like a stable rotation.  The discrepancy says that the $2d_{5/2}$
 orbit must not so much join in the quadrupole correlations
 (or the $2d_{5/2}$ orbit must be far from $1g_{9/2}$) and hence
 the deformation is not large for $^{90}$Ru, which is consistent
 with a smaller deformation (0.16) adopted in Ref. \cite{Marg2,Sun2}.

    As mentioned in Ref. \cite{Hase1}, the $QQ$ force gives inverse magnitudes
 to the interaction matrix elements
 $ \langle (g_{9/2})^2_{J,T=1} |H^{\tau =0}_{QQ}| (g_{9/2})^2_{J,T=1} \rangle $
 with $J=6$ and $J=8$ contrary to those of the ordinary effective interaction
 \cite{Serduke}.  This defect has a bad influence on the $J-\omega$ graph at $J=8$.
 If we replace the $J=6$ and $J=8$ matrix elements with those of Ref.
 \cite{Serduke}, the slight backbending at $J=8$ in the calculated result
 for $^{88}$Ru disappears as shown by the dot line (Y) in Fig. \ref{fig7}.
 Figures \ref{fig1} to \ref{fig6} are not free from the same influence, either.
 This defect, however, does not change the general situation.

\begin{table}[b]
\caption{$B(E2:J_i \rightarrow J_f)$ for the yrast states of $^{88}$Ru and $^{90}$Ru
         calculated with the different strengths of the $QQ$ force, A, B, C, D and E.
         The last column X shows the values obtained in the model space
         $(2p_{1/2},1g_{9/2},2d_{5/2})$.}
\begin{tabular}{c|cccccc}   \hline
  & \multicolumn{6}{c}{$B(E2:J_i \rightarrow J_f)$ in $e^2$fm$^4$} \\
  $J_i \rightarrow J_f$ & A   &  B  &  C  &  D  &  E  & X   \\ \hline
   $^{88}$Ru            &     &     &     &     &     &     \\
  $2 \rightarrow 0$     & 460 & 543 & 536 & 522 & 612 & 510 \\
  $4 \rightarrow 2$     & 630 & 730 & 722 & 704 & 832 & 730 \\
  $6 \rightarrow 4$     & 672 & 790 & 780 & 758 & 914 & 582 \\
  $8 \rightarrow 6$     & 697 & 847 & 833 & 806 & 964 & 601 \\
  $10 \rightarrow 8$    & 771 & 897 & 884 & 862 & 999 & 619 \\
  $12 \rightarrow 10$   & 775 & 896 & 882 & 859 & 999 & 568 \\
  $14 \rightarrow 12$   & 748 & 880 & 866 & 844 & 982 & 532 \\
  $16 \rightarrow 14$   & 671 & 853 & 839 & 814 & 962 & 489 \\
  $18 \rightarrow 16$   & 622 & 820 & 806 & 782 & 931 & 430 \\
  $20 \rightarrow 18$   &  71 & 760 & 748 & 725 & 872 & 348 \\ \hline

  $^{90}$Ru             &     &     &     &     &     &     \\
  $2 \rightarrow 0$     & 296 & 339 & 348 & 339 & 482 & 567 \\
  $4 \rightarrow 2$     & 385 & 451 & 466 & 456 & 663 & 793 \\
  $6 \rightarrow 4$     & 294 & 388 & 418 & 418 & 698 & 713 \\
  $8 \rightarrow 6$     & 235 & 236 & 213 & 234 & 654 & 657 \\
  $10 \rightarrow 8$    & 309 & 322 & 325 & 331 & 764 & 673 \\
  $12 \rightarrow 10$   & 284 & 290 & 289 & 270 & 501 & 621 \\
  $14 \rightarrow 12$   & 257 & 265 & 227 & 218 & 600 & 538 \\
  $16 \rightarrow 14$   & 238 & 274 & 296 & 299 & 396 & 440 \\
  $18 \rightarrow 16$   & 252 & 275 & 296 & 300 & 597 & 413 \\
  $20 \rightarrow 18$   & 191 & 199 & 202 & 205 & 542 & 335 \\ \hline
\end{tabular}
\label{table3}
\end{table}

\subsection{$B(E2)$ and $Q$ moment}

   We have discussed the energy levels, $J-\omega$ graph and
 $\langle n_a \rangle$ so far.
 The $B(E2)$ value and $Q$ moment are good physical quantities to see
 the characteristics of the quadrupole correlations.  We calculated
 the $B(E2)$ values and $Q$ moments for the yrast states of $^{88}$Ru
 and $^{90}$Ru, using the different strengths of the $QQ$ force A, B, C, D
 and E (corresponding to those in Figs. \ref{fig3} - \ref{fig6}) and
 the model X (corresponding to X in Fig. \ref{fig7}).
  We used the effective charges $e_p=1.5e$ and $e_n=0.5e$,
 to compare the relative values of electric quadrupole quantities
 obtained with the different strengths of the $QQ$ force.
 The calculated results are tabulated in Tables \ref{table3}
 and \ref{table4}.

  In Table \ref{table3}, the calculated $B(E2)$ values of $^{88}$Ru are
 much larger than those of $^{90}$Ru,  providing that the energy levels
 of both nuclei are approximately reproduced.  The ratios of the $B(E2)$
 values are more than 1.6 in the calculation A.  In other words,
 the quadrupole correlations are much more enhanced in the $N=Z$ nucleus
 $^{88}$Ru than in $^{90}$Ru.  This is consistent with the fact
 that a larger deformation is employed for $^{88}$Ru as compared with
 $^{90}$Ru in the projected shell model \cite{Marg2,Sun2}.
 
\begin{table}[b]
\caption{Quadrupole moment $Q(J)$ in $e\cdot$fm$^2$ for the yrast states
         of $^{88}$Ru and $^{90}$Ru calculated with the different strengths
         of the $QQ$ force, A, B, C, D and E.}
\begin{tabular}{cc|rrrrr}   \hline
 & $ J $     & A     &  B    &  C    &  D    &  E     \\ \hline
$^{88}$Ru
 & $ 2 $     &  1.4  &  8.9  &  8.1  &  6.3  & 28.7   \\
 & $ 4 $     & -1.9  &  8.1  &  7.5  &  5.1  & 31.0   \\
 & $ 6 $     & 20.1  & 28.3  & 27.6  & 25.7  & 41.7   \\
 & $ 8 $     & 23.6  & 30.8  & 30.2  & 28.6  & 41.6   \\
 & $10 $     & 26.6  & 33.3  & 32.7  & 31.3  & 42.1   \\
 & $12 $     & 31.1  & 36.5  & 36.3  & 35.2  & 43.6   \\
 & $14 $     & 32.2  & 37.5  & 37.2  & 36.3  & 43.5   \\
 & $16 $     & 26.9  & 37.6  & 37.1  & 35.8  & 43.6   \\
 & $18 $     & 26.6  & 39.5  & 38.9  & 37.6  & 45.2   \\
 & $20 $     &-14.9  & 43.1  & 42.5  & 41.5  & 47.8   \\ \hline

$^{90}$Ru
 & $ 2 $     & -12.0 & -16.4 & -16.7 & -15.8 & -14.4  \\
 & $ 4 $     & -16.6 & -21.6 & -22.5 & -21.5 & -12.8  \\
 & $ 6 $     &  12.5 &   5.1 &   4.7 &   3.7 &  15.9  \\
 & $ 8 $     &  15.5 &  14.3 &  16.5 &  16.9 &  36.0  \\
 & $10 $     &  11.5 &   9.9 &  11.7 &  12.2 &  38.8  \\
 & $12 $     &   5.6 &   5.0 &   5.6 &   5.7 &  53.7  \\
 & $14 $     &  -3.6 &   4.2 &   6.8 &   7.7 &  68.4  \\
 & $16 $     &   1.8 &   4.1 &   7.4 &   8.0 &  36.2  \\
 & $18 $     &  -0.7 &   0.9 &   3.6 &   4.5 &  33.5  \\
 & $20 $     &  -4.1 &  -3.9 &  -2.7 &  -2.0 &  35.7  \\ \hline

\end{tabular}
\label{table4}
\end{table}

 The modifications of the $QQ$ interaction, B, C and D, somewhat enlarge
 the $B(E2)$ values both for $^{88}$Ru and $^{90}$Ru. The ratios of
 the $B(E2)$ values for $^{88}$Ru to those for $^{90}$Ru are still large. 
 The very strengthened $QQ$ force E, which is required to reproduce
 the slope of the $J-\omega$ graph for $^{88}$Ru, enlarges the $B(E2)$ values
 fairly for $^{88}$Ru and drastically for $^{90}$Ru.
  We have already seen that the strengthened $QQ$ force E ruins the pattern of
 energy levels for $^{90}$Ru.  The enhanced $B(E2)$ values in the column E of
 Table \ref{table3} are therefore too large for $^{90}$Ru. 
 We do not adopt the large $B(E2)$ values in the column X for $^{90}$Ru
 from the same reason, either.
  The quadrupole correlations must not be enhanced too much and
 the contribution of the $2d_{5/2}$ orbit should be small for $^{90}$Ru.
 Namely, $^{90}$Ru may not be largely deformed.
  The truncated configuration space $(2p_{1/2},1g_{9/2},2d_{5/2})$ yields
 $B(E2)$ values comparable to those of the result A for $^{88}$Ru, in spite of
 the small space.  The expansion of the model space by adding 
 $2d_{5/2}$ to $(2p_{3/2},1f_{5/2},2p_{1/2},1g_{9/2})$ can make the $B(E2)$
 values larger, which could be appropriate to the enhanced
 quadrupole correlations in $^{88}$Ru.

   The calculated quadrupole moments $Q(J)$ tabulated in Table \ref{table4}
 show the same results as the $B(E2)$ values.
  From Table \ref{table4}, we can say as follows:
 The small enhancements of the $p-n$ $QQ$ interaction (B and C) change
 insignificantly the structure of $^{88}$Ru and $^{90}$Ru,
 while the strong enhancement of the $QQ$ force by 1.5 times (E) changes
 the structure of $^{88}$Ru drastically.  The large and roughly constant
 $Q$ moments of $^{88}$Ru suggest the quadrupole deformation.
 If the result E should not be adopted for $^{90}$Ru, Table \ref{table4} and
 the energy levels insist that $^{90}$Ru does not have a large deformation.

   The calculated $B(E2)$ values and $Q$ moments in Tables \ref{table3} and
 \ref{table4} testify the structural change due to the particle pair alignment
 at $8^+$ in $^{90}$Ru, in contrast to $^{88}$Ru.
 The $B(E2: 8^+ \rightarrow 6^+)$ value decreases and $Q$ moment increases
 at $8^+$ in $^{90}$Ru, while the two values do not show any abrupt changes
 at $8^+$ in $^{88}$Ru.  On the other hand, the simultaneous alignments
 of proton and neutron pairs at $16^+$ leave a sign in the $B(E2)$ values
 and $Q$ moment in the result A both for $^{88}$Ru and $^{90}$Ru.

\section{Prediction for $^{89}$Ru}

\begin{figure}[b]
\includegraphics[width=7.5cm,height=7.5cm]{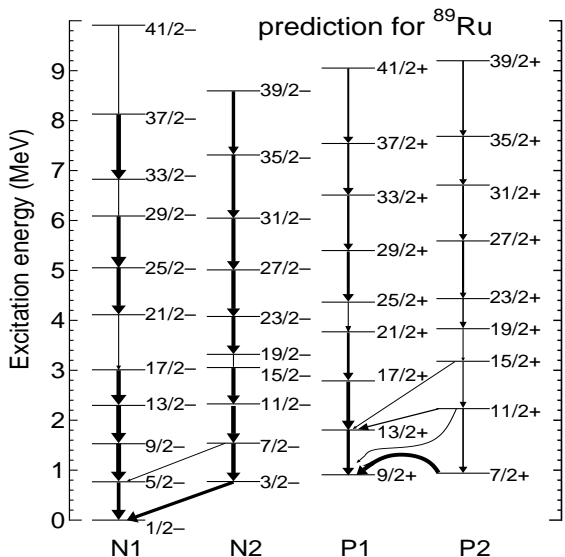}
  \caption{Energy levels predicted for $^{89}$Ru.  The widths of the arrows
  show the relative $B(E2)$ values.}
  \label{fig8}
\end{figure}

\begin{table}[b]
\caption{Expectation values of proton and neutron numbers
         in the respective orbits for the yrast states of $^{89}$Ru
         calculated with the $QQ$ force strength A.
         Calculated $Q$ moments are also tabulated.}
\begin{tabular}{r|cccc|cccc|r}   \hline
       & \multicolumn{4}{c}{proton} & \multicolumn{4}{|c|}{neutron} & \\
   $J^\pi$ & $p_{3/2}$ & $f_{5/2}$ & $p_{1/2}$ & $g_{9/2}$
       & $p_{3/2}$ & $f_{5/2}$ & $p_{1/2}$ & $g_{9/2}$ & $Q$  \\ \hline
 $1/2^-$ & 3.89 & 5.79 & 0.67 & 5.64 & 3.93 & 5.89 & 1.22 & 5.95 & -0.005 \\
 $3/2^-$ &  3.88 & 5.79 & 0.63 & 5.69 & 3.92 & 5.90 & 1.22 & 5.96 & -9.1 \\
 $5/2^-$ & 3.88 & 5.79 & 0.64 & 5.69 & 3.94 & 5.88 & 1.22 & 5.96 & -11.7 \\
 $7/2^-$ &  3.88 & 5.79 & 0.63 & 5.71 & 3.92 & 5.90 & 1.21 & 5.96 & -17.0 \\
 $9/2^-$ & 3.88 & 5.78 & 0.63 & 5.70 & 3.94 & 5.88 & 1.22 & 5.96 & -16.8 \\
$11/2^-$ &  3.89 & 5.79 & 0.72 & 5.60 & 3.92 & 5.90 & 1.21 & 5.97 & 8.6 \\
$13/2^-$ & 3.89 & 5.79 & 0.73 & 5.59 & 3.93 & 5.87 & 1.22 & 5.97 & 10.6 \\
$15/2_2^-$ &  3.91 & 5.82 & 0.91 & 5.37 & 3.93 & 5.90 & 1.21 & 5.96 & 17.7 \\
$17/2^-$ & 3.91 & 5.81 & 0.90 & 5.39 & 3.94 & 5.88 & 1.24 & 5.94 & 18.6 \\
$19/2^-$ &  3.94 & 5.89 & 1.09 & 5.08 & 3.98 & 5.95 & 1.78 & 5.29 & 5.1 \\
$21/2^-$ & 3.93 & 5.91 & 1.09 & 5.06 & 3.98 & 5.96 & 1.83 & 5.24 & 5.9 \\
$23/2^-$ &  3.95 & 5.89 & 1.10 & 5.06 & 3.98 & 5.95 & 1.82 & 5.24 & 5.9 \\
$25/2^-$ & 3.94 & 5.92 & 1.09 & 5.05 & 3.98 & 5.97 & 1.87 & 5.18 & 5.5 \\
$27/2^-$ &  3.95 & 5.89 & 1.11 & 5.05 & 3.98 & 5.96 & 1.86 & 5.19 & 6.0 \\
$29/2^-$ & 3.94 & 5.93 & 1.09 & 5.04 & 3.99 & 5.97 & 1.91 & 5.13 & 4.6 \\
$31/2^-$ &  3.96 & 5.90 & 1.11 & 5.04 & 3.99 & 5.97 & 1.90 & 5.14 & 5.2 \\
$33/2^-$ & 3.90 & 5.80 & 0.75 & 5.55 & 3.95 & 5.88 & 1.25 & 5.92 & 20.3 \\
$35/2^-$ &  3.96 & 5.90 & 1.12 & 5.02 & 3.99 & 5.99 & 1.97 & 5.06 & 4.7 \\
$37/2^-$ & 3.91 & 5.81 & 0.82 & 5.47 & 3.96 & 5.88 & 1.26 & 5.90 & 19.9 \\
$39/2^-$ &  3.97 & 5.91 & 1.10 & 5.02 & 3.99 & 5.99 & 1.98 & 5.04 & 4.3 \\
$41/2^-$ & 3.96 & 5.96 & 1.06 & 5.02 & 3.99 & 5.99 & 1.96 & 5.06 & 3.3 \\ \hline

 $7/2^+$ &  3.93 & 5.85 & 1.13 & 5.09 & 3.94 & 5.89 & 1.41 & 5.75 & 19.0 \\
 $9/2^+$ & 3.94 & 5.86 & 1.22 & 4.98 & 3.95 & 5.90 & 1.48 & 5.66 & 13.1 \\
$11/2^+$ &  3.96 & 5.90 & 1.57 & 4.57 & 3.98 & 5.95 & 1.83 & 5.25 & 19.0 \\
$13/2^+$ & 3.94 & 5.87 & 1.31 & 4.88 & 3.96 & 5.92 & 1.61 & 5.51 & 11.8 \\
$15/2^+$ &  3.95 & 5.91 & 1.46 & 4.68 & 3.96 & 5.92 & 1.55 & 5.57 & 16.7 \\
$17/2^+$ & 3.95 & 5.89 & 1.43 & 4.72 & 3.97 & 5.94 & 1.74 & 5.34 & 11.7 \\
$19/2^+$ &  3.94 & 5.90 & 1.39 & 4.76 & 3.96 & 5.92 & 1.59 & 5.53 & 2.4 \\
$21/2^+$ & 3.98 & 5.95 & 1.77 & 4.30 & 3.98 & 5.97 & 1.88 & 5.17 & 9.7 \\
$23/2^+$ &  3.97 & 5.94 & 1.69 & 4.40 & 3.98 & 5.95 & 1.75 & 5.32 & -5.5 \\
$25/2^+$ & 3.96 & 5.91 & 1.49 & 4.64 & 3.97 & 5.93 & 1.68 & 5.41 & 3.3 \\
$27/2^+$ &  3.97 & 5.95 & 1.70 & 4.38 & 3.98 & 5.96 & 1.84 & 5.21 & -2.8 \\
$29/2^+$ & 3.96 & 5.93 & 1.62 & 4.48 & 3.98 & 5.96 & 1.81 & 5.25 & 0.5 \\
$31/2_2^+$ &  3.98 & 5.96 & 1.77 & 4.29 & 3.99 & 5.98 & 1.93 & 5.10 & -3.0 \\
$33/2^+$ & 3.98 & 5.95 & 1.75 & 4.32 & 3.99 & 5.98 & 1.93 & 5.10 & -3.5 \\
$35/2^+$ &  3.98 & 5.96 & 1.78 & 4.27 & 3.99 & 5.98 & 1.92 & 5.11 & -1.7 \\
$37/2^+$ & 3.98 & 5.96 & 1.78 & 4.28 & 3.99 & 5.98 & 1.93 & 5.10 & -4.8 \\
$39/2^+$ &  3.99 & 5.98 & 1.89 & 4.15 & 3.99 & 5.99 & 1.95 & 5.07 & -5.5 \\
$41/2^+$ & 3.99 & 5.98 & 1.89 & 4.15 & 3.99 & 5.99 & 1.96 & 5.07 & -7.9 \\ \hline

\end{tabular}
\label{table5}
\end{table}

  The $^{89}$Ru isotope between $^{88}$Ru and $^{90}$Ru has not experimentally
 been observed yet.  Our model, however, predicts interesting features of
 $^{89}$Ru.
   Figure \ref{fig8} shows the energy levels and relative $B(E2)$ values
 obtained using the parameter set A for $^{89}$Ru.  The collective states
 connected by large $B(E2)$ values are divided into four bands.
   They are the yrast states except for
 $15/2^-$ and $31/2^+$.  Exceptionally, we select the second states
 $(15/2)_2^-$ and $(31/2)_2^+$ as collective states  based on the $B(E2)$ values
 and $Q$ moments, which are adjacent to the yrast states $(15/2)_1^-$
 and $(31/2)_1^+$ respectively. 
 The relative $B(E2)$ values are denoted by the widths of the arrows in Fig.
 \ref{fig8}.  The inter-band $E2$ transitions which are not shown in Fig.
 \ref{fig8} are weak.
 
   It is remarkable that the predicted ground state of the middle
 $1g_{9/2}$-subshell nucleus $^{89}$Ru is the $1/2^-$ state.
  This extraordinary event is reasonable from the systematic lowering
 of the $1/2^-$ state with decreasing $N$ in odd-$A$ Ru isotopes
 as seen in Fig. \ref{fig2}.  Our model reproduces the systematic
 behavior of $1/2^-$.
 Look at the expectation values of proton and neutron numbers 
 $\langle n_a \rangle$ for the band-head states $1/2^-$, $3/2^-$, $9/2^+$
 and $7/2^+$ of $^{89}$Ru which are tabulated in Table \ref{table5}.
 This table shows that the $1/2^-$ state has more protons in $1g_{9/2}$
 than the $9/2^+$ state.
  From the comparison of Table \ref{table5} with Tables \ref{table1} and
 \ref{table2}, the $1/2^-$ state resembles the ground state of $^{88}$Ru
 and the $9/2^+$ state resembles the ground state of $^{90}$Ru.
   Roughly speaking, the $1/2^-$ state is constructed
 by adding one neutron to $^{88}$Ru($0^+$), and the $9/2^+$ state by removing
 one neutron from $^{90}$Ru($0^+$).
   The $B(E2)$ values of the negative parity bands larger than those of
 the positive parity bands indicate the stronger collectivity of
 the negative parity bands.  This corresponds to the result shown
 in Table \ref{table3} that the $B(E2)$ values of $^{88}$Ru are larger
 than those of $^{90}$Ru.
 From these comparisons, the difference between the $1/2^-$ state and
 the $9/2^+$ state can be understood in terms of the $\alpha$-like $2p-2n$
 correlations mentioned in the interpretation of the difference
 between $^{88}$Ru and $^{90}$Ru.
  The strong $\alpha$-like $2p-2n$ correlations pull up more protons to $1g_{9/2}$
 in $1/2^-$ than in $9/2^+$, because the disturbing extra neutron is absent
 in $1g_{9/2}$ for the $1/2^-$ state.
  The inversion of $9/2^+$ and $1/2^-$ says that the $\alpha$-like $2p-2n$
 correlations give a larger energy gain to the $1/2^-$ state and the larger
 correlation energy compensates the energy loss of more nucleon jumps
 to the $1g_{9/2}$ subshell in $1/2^-$ .
 
\begin{figure}
\includegraphics[width=4.9cm,height=6.0cm]{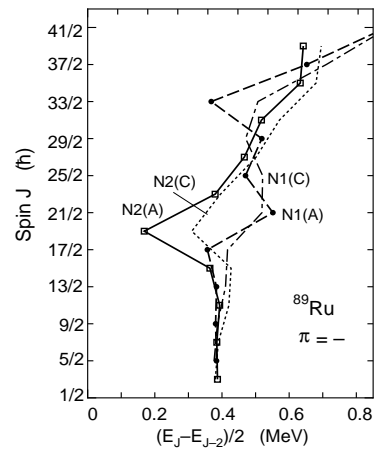}
  \caption{The $J-\omega$ graph for the negative parity bands N1 and N2
           of $^{89}$Ru.
           The labels A and C stand for the parameter sets A and C.}
  \label{fig9}
\end{figure}

\begin{figure}
\includegraphics[width=4.9cm,height=6.0cm]{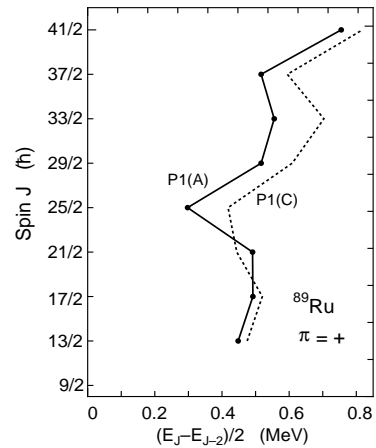}
  \caption{The $J-\omega$ graph for the positive parity band P1 of $^{89}$Ru.}
  \label{fig10}
\end{figure}

  We can expect that the $\Delta J=2$ bands on the $1/2^-$ and $9/2^+$
 states are similar to the ground-state bands of $^{88}$Ru and
 $^{90}$Ru, respectively.  In fact, the $J-\omega$ graphs for the two bands
 N1 and P1 of $^{89}$Ru which are shown in Figs. \ref{fig9} and \ref{fig10}
 are similar to those of $^{88}$Ru and $^{90}$Ru in Figs. \ref{fig5} and
 \ref{fig6}.    Figure \ref{fig9} suggests no backbending
 at $17/2^-$ ($1/2^- +8$) in the negative parity band N1,
 while Fig. \ref{fig10} predicts a backbending phenomenon at $25/2^+$
 ($9/2^+ +8$) in the positive parity band P1.
 The backbending at $25/2^+$ in the band P1 seems to be caused
 by the proton pair alignment parallel to the spin of the last odd neutron
 in $1g_{9/2}$, corresponding to the neutron pair alignment at $8^+$ in $^{90}$Ru.
 The increase of the proton number $\langle n_{g9/2} \rangle$ at $25/2^+$
 testifies the proton pair alignment in $1g_{9/2}$. The small $B(E2)$ value
 from $25/2^+$ to $21/2^+$ and the decrease of the calculated $Q$ moment
 at $25/2^+$ show the structural change there.
   The coincident increase of the neutron number $\langle n_{g9/2} \rangle$
 at $25/2^+$ gives another evidence of strong $p-n$ correlations in $1g_{9/2}$.
 For the negative parity band N1, the value of
 $B(E2:17/2^- \rightarrow 13/2^-)$ does not show any sign of
 such a structural change and the expectation values of proton and neutron
 numbers show no abrupt change, which corresponds to no backbending at $8^+$
 in $^{88}$Ru. 
 
   Figure \ref{fig9}, however, predicts backbending at $33/2^-$ ($1/2^- +16$)
 in the band N1. The simultaneous increases of proton and neutron numbers
 $\langle n_{g9/2} \rangle$ at $33/2^-$ (see Table \ref{table5}) say that
 this backbending is due to the simultaneous alignments of proton and neutron
 pairs ($J=16$) in $1g_{9/2}$, corresponding to the four nucleon alignment
 in $^{88}$Ru.  The small value of $B(E2:33/2^- \rightarrow 29/2^-)$ and
 the increase of the $Q$ moment at $33/2^-$ testify the structure change.
   Figure \ref{fig10} shows a sign of another backbending at $37/2^+$
 in the band P1, that is possibly the alignment of two protons and three
 neutrons in $1g_{9/2}$, corresponding to the alignment at $16^+$ in $^{90}$Ru.
 The spin of the last odd neutron is parallel to the spin of rotation or
 alignment in the band P1.
 

   Figure \ref{fig9} shows also the $J-\omega$ graph for the negative
 parity band N2.  This figure with Fig. \ref{fig8} says that the low-lying
 collective states $3/2^-$, $7/2^-$, $11/2^-$ and $(15/2)_2^-$ of the band N2
 and $5/2^-$, $9/2^-$, $13/2^-$ and $17/2^-$ of the band N1 are the partners
 in the angular momentum coupling $1/2^- \otimes J$ ($=J \pm 1/2$).
 The similar $B(E2)$ values and similar $Q$ moments support the picture.
 The most remarkable backbending in $^{89}$Ru takes place at the $19/2^-$ state
 of the negative parity band N2.  The small $B(E2)$ values
 from $19/2^-$ to $(15/2)_1^-$, $(15/2)_2^-$ and $17/2^-$ indicate
 a clear structure change at $19/2^-$.
 The $Q$ moment decreases abruptly at $19/2^-$. 
  For the $J \ge 19/2^-$ states of the band N2, the $Q$ moments are
 nearly constant and so is the slope of $J-\omega$ graph.
 This phenomenon cannot be explained by the nucleon pair alignment coupled to
 $J=8$ ($T=1$), because the $J=19/2$ state cannot be constructed by the coupling
 $1/2 \otimes 8$.  If the structure change is due to a kind of alignment,
 the phenomenon is attributed to the $p-n$ alignment $J=9$ ($T=0$).
 
 In $^{89}$Ru, the efficient way to construct the $J=9$ $p-n$ pair is
 one proton jump to $1g_{9/2}$.  When one $p-n$ pair aligns to $J=9$ ($T=0$)
 in $1g_{9/2}$, another pair which breaks away from the $\alpha$-like $2p-2n$
 correlations is still possible to join in the monopole ($J=0$, $T=1$) pairing
 correlations, and to couple with the last odd nucleon in $2p_{1/2}$
 to the total isospin $T=1/2$.
 The decreases of neutron and proton numbers $\langle n_{g9/2} \rangle$
 at $19/2^-$ testify the decline of the $\alpha$-like $2p-2n$ correlations
 due to the breaking away of the $J=9$ $p-n$ pair from a $T=0$ $2p-2n$ quartet.
 It should be noted that the disunion of a $T=0$ $2p-2n$ quartet to the $T=0$
 and $T=1$ pairs is prohibited for even-even nuclei.
  The $T=0$ $p-n$ alignment at $19/2^-$ seems to be a unique phenomenon in
 the $1g_{9/2}$-subshell odd-$A$ nuclei with $N=Z \pm 1$ such as $^{89}$Ru.
 (The $T=0$ $p-n$ alignment could take place in $N=Z$ odd-odd nuclei.)
 In this connection, the small $E2$ values from $21/2^-$ to $17/2^-$
 is notable.  It suggests that the $p-n$ alignment ($J=9$, $T=0$)
 contributes to the $J \ge 21/2^-$ states of the band N1.
  Actually, the states $21/2^-$, $25/2^-$ and $29/2^-$ of the band N1
 resemble the states $23/2^-$, $27/2^-$ and $31/2^-$ of the band N2,
 with respect to the energy levels, expectation values of nucleon numbers
 $\langle n_a \rangle$, $B(E2)$ values and $Q$ moments.
 They could be members of the collective excitations coupled with
 the three nucleons $2p_{1/2}^\pi(1g_{9/2}^\pi 1g_{9/2}^\nu)_{J=9,T=0}$.
 
  The positive parity band P2 shows a rather complicated behavior.  
 The very low-lying $7/2^+$ state is apparently related to the state
 of three nucleons with $J=j-1$ in a high-spin orbit $j$ \cite{Bohr}.
 The small $B(E2:19/2^+ \rightarrow 15/2^+)$ value and the abrupt decrease
 of the $Q$ moment at $19/2^+$ testify a structure change at the $19/2^+$
 state.

\section{Conclusions}

  We have carried out the shell model calculations on the spherical basis
 using the extended $P+QQ$ Hamiltonian with a single set of parameters
 in the model space $(2p_{3/2},1f_{5/2},2p_{1/2},1g_{9/2})$.
 The calculations reproduce qualitatively well the overall energy levels
 observed in the Ru isotopes, $^{88}$Ru, $^{90}$Ru, $^{91}$Ru, $^{92}$Ru,
 $^{93}$Ru and $^{94}$Ru.  The extended $P+QQ$ model is confirmed to be useful
 in the heaviest $N \approx Z$ nuclei.  The results testify the enhancement
 of the quadrupole correlations at the $N=Z$ nucleus $^{88}$Ru as compared with
 the other Ru isotopes.

  However, the disagreement between theory and experiment for $^{88}$Ru cannot be
 disregarded.  The slope of the $J-\omega$ graph showing the moment of inertia
 and the durable increase of angular frequency are not sufficiently reproduced
 for $^{88}$Ru with the $QQ$ force strength commonly fixed to all the Ru isotopes.
  The theoretical analysis suggests a further enhancement of the quadrupole
 correlations, and recommends us to use a stronger $QQ$ force for $^{88}$Ru.
 We have tried to strengthen the $p-n$ $QQ$ interaction in the two ways
 so as to conserve and not to conserve the isospin of eigenstates,
 and also to strengthen all the $p-p$, $n-n$ and $p-n$ parts of the isoscalar
 $QQ$ force.  Within a small enhancement, however, there is little to choose
 between them in the present calculations.
   Anyway, the present study indicates a special enhancement of the quadrupole
 correlations in the $N=Z$ nucleus $^{88}$Ru.
  This is consistent with the large deformation of $^{88}$Ru in contrast to
 $^{90}$Ru which is predicted by the projected shell model calculation
 on the deformed basis \cite{Marg2,Sun2}.

   The requirement of the enhanced $QQ$ force for $^{88}$Ru possibly means
 that the configuration space should be extended in our spherical shell model.
 We have investigated the contribution of the $2d_{5/2}$ orbit which is
 expected to mix with the $1g_{9/2}$ orbit through the large $Q$ matrix element.
 The truncated space $(2p_{1/2},1g_{9/2},2d_{5/2})$ can easily reproduce
 the slope of $J-\omega$ graph observed in $^{88}$Ru.  The result suggests
 that the $2d_{5/2}$ orbit contributes to the quadrupole correlations,
 which supports that $^{88}$Ru is deformed.
  Contrary to this, the same calculation requires a much smaller
 contribution of the $2d_{5/2}$ orbit to $^{90}$Ru.  It is, therefore,
 likely that $^{88}$Ru is deformed while $^{90}$Ru is not largely deformed
 as known from the observed $J-\omega$ graphs.
   This situation still demands different $QQ$ force strengths for $^{88}$Ru 
 and $^{90}$Ru in our spherical shell model calculation.  There is not
 a selfconsistent way to determine the $QQ$ force strength. 
 An additional constraint, for instance with respect to the $Q$ moment value,
 is necessary for it.  The condition is the same for the treatment
 on the deformed basis \cite{Marg2,Sun2}.  The deformation should be
 selfconsistently determined there.

    Our model with a single set of parameters, however, is capable of
 describing the difference between $^{88}$Ru and $^{90}$Ru.
 The calculations have presented a useful knowledge of the structure
 of Ru isotopes.
  The contrast features of $^{88}$Ru and $^{90}$Ru owe to the $\alpha$-like
 ($T=0$) $2p-2n$ correlations depending on the shell structure
 in the $1g_{9/2}$ subshell nuclei.
  In the $^{90}$Ru isotope with one extra neutron pair which does not
 join in the $\alpha$-like $2p-2n$ correlations, the extra neutron pair
 aligns easily to $J=9/2+7/2=8$ ($T=1$) in $1g_{9/2}$.  In contrast to this,
 the $\alpha$-like ($T=0$) $2p-2n$ correlations hinder the single
 nucleon-pair alignment coupled to $J=8$ ($T=1$) till the simultaneous
 alignments of proton and neutron pairs at $J=2 \times 8$ ($T=0$),
 in the $N=Z$ even-even nucleus $^{88}$Ru.
 
  The shell structure produces characteristic bands with opposite parities
 in $^{89}$Ru.  The following predictions are obtained for $^{89}$Ru:
 The $1/2^-$ state is the ground state.  There are three characteristic bands.
 The negative parity band N1 on $1/2^-$, which resembles the ground-state band
 of $^{88}$Ru, shows backbending at $33/2^-$ caused by the simultaneous
 alignments of proton and neutron pairs coupled to $J=16$ in $1g_{9/2}$.
 The $J \le 15/2^-$ states of another negative parity band N2 on $3/2^-$ are
 the partners of the $J \le 17/2^-$ states of the band N1.
  The band N2 shows a unique backbending at $19/2^-$ caused by the $p-n$ pair
 alignment coupled to $J=9$ ($T=0$) in $1g_{9/2}$.
 The positive parity band P1 on $9/2^+$, which resembles the ground-state band
 of $^{90}$Ru, displays backbending due to the proton pair alignment
 $J=8$ ($T=1$) parallel to the spin of the last odd neutron in $1g_{9/2}$.
 These predictions wait for experimental examinations.



\end{document}